\documentclass[10pt,conference]{IEEEtran}

\usepackage{amssymb}
\usepackage{amsmath}
\usepackage{cite}
\usepackage{url}
\usepackage{xcolor}
\usepackage{cite,graphicx,amsmath,amssymb}
\usepackage{fancyhdr}
\usepackage{mdwmath}
\usepackage{mdwtab}
\usepackage{caption}
\usepackage{amsthm}
\usepackage{setspace}
\usepackage{algorithm}
\usepackage{algorithmic}
\usepackage{pdfpages}
\usepackage{mathtools}
\usepackage{subcaption}
\usepackage{bbm}

\graphicspath{{./src/img/}}

\newtheorem{theorem}{Theorem}

\newtheorem{lemma}{Lemma}

\newtheorem{corollary}{Corollary}

\allowdisplaybreaks
\expandafter\def\expandafter\normalsize\expandafter{%
    \normalsize%
    \setlength\abovedisplayskip{4pt}%
    \setlength\belowdisplayskip{4pt}%
    \setlength\abovedisplayshortskip{2pt}%
    \setlength\belowdisplayshortskip{2pt}%
}
\title{Performance Bounds of Near-Field Sensing with Circular Arrays}
\author{
    \IEEEauthorblockN{Zhaolin Wang\IEEEauthorrefmark{1}, Xidong Mu\IEEEauthorrefmark{1}, and Yuanwei Liu\IEEEauthorrefmark{1}\\
\IEEEauthorblockA{\IEEEauthorrefmark{1}Queen Mary University of London, London, UK.
\\E-mail: \{zhaolin.wang, xidong.mu, yuanwei.liu\}@qmul.ac.uk}}
\vspace{-0.7cm}
}

\begin{document}

\maketitle
\begin{abstract}
    The performance bounds of near-field sensing are studied for circular arrays, focusing on the impact of bandwidth and array size. The closed-form Cramér-Rao bounds (CRBs) for angle and distance estimation are derived, revealing the scaling laws of the CRBs with bandwidth and array size. Contrary to expectations, enlarging array size does not always enhance sensing performance. Furthermore, the asymptotic CRBs are analyzed under different conditions, unveiling that the derived expressions include the existing results as special cases. Finally, the derived expressions are validated through numerical results.
\end{abstract}

\vspace{-0.3cm}
\section{Introduction}
Multiple-input multiple-output (MIMO) has played a critical role in wireless sensing systems \cite{fishler2004mimo} and has also been regarded as a key enabling technology for integrating wireless sensing functionalities into the mobile communication networks \cite{liu2022integrated}. The current fifth-generation (5G) wireless network is based on the massive MIMO concept, which was first introduced by Marzetta in 2010 \cite{5595728}. A typical 5G massive MIMO base station (BS) would be equipped with 64 antennas. Looking forward to the sixth-generation (6G) era in the 2030s, the new concepts of extremely large-scale MIMO (XL-MIMO) and extremely large-aperture antenna array (ELAA), which may include hundreds or even thousands of antennas, have been proposed and received growing attention \cite{wang2024tutorial}. This will not only increase the aperture size of antenna arrays but also bring about fundamental changes to the electromagnetic properties of wireless signals \cite{liu2023near_tutorial, 10380596, 10639537}. In particular, as the array aperture grows, the near-field region around BSs can be greatly expanded. Such expansion can be more significant in the high-frequency bands, such as millimeter wave and terahertz bands. Generally speaking, in the near-field region, the spherical-wave propagation of wireless signals is dominant, which is fundamentally different from the planar-wave propagation observed in far-field scenarios. Therefore, it is crucial to reevaluate the wireless sensing performance from a near-field perspective. 

There are two key factors determining wireless sensing performance: \emph{bandwidth} and \emph{array size}. In conventional far-field sensing, these factors are primarily associated with the resolution of \emph{distance} and \emph{angle} estimation, respectively \cite{guerci2015joint}. However, in near-field sensing, the array size not only affects the resolution of angle estimation but also impacts distance estimation due to spherical-wave propagation, thus alleviating the stringent requirement on bandwidth. This unique advantage motivates extensive studies on near-field sensing performance in narrowband systems, disregarding the influence of bandwidth \cite{el2010conditional, 9439203, wang2023near, wang2023cram}. Additionally, while most existing studies focus on uniform linear arrays (ULAs), the effective array aperture of ULAs significantly diminishes near the edges, which can severely impair sensing performance at large incident and departure angles. Circular arrays have been proven to be a promising solution to address this challenge due to their isotropic radiation properties \cite{wu2023enabling, wang2023rethinking}.  

Against the above background, this paper studies the joint impact of bandwidth and array size on near-field sensing performance with circular arrays. The analysis is based on the widely exploited Cramér-Rao bound (CRB) framework \cite{el2010conditional, 9439203, wang2023near, wang2023cram} and the most popular orthogonal frequency-division multiplexing (OFDM) wideband signaling method \cite{sturm2011waveform}. Based on these preconditions, we derive closed-form expressions for the CRBs of angle and distance estimation, which unveil the joint effects of bandwidth and array size on near-field sensing performance. We further explore the asymptotic behavior of these CRBs under various scenarios to elucidate the connections between our new findings and existing literature. Finally, numerical results are provided to validate the analytical results.

\section{System Model} \label{sec:system_model}
We study near-field sensing in a legacy wideband OFDM system with an $N$-antenna base station (BS). The BS carries out mono-static sensing for a point target located within the near-field region of the BS. We assume a shared uniform circular array (UCA) for transmitting and receiving at the BS through the use of circulators and the perfect self-interference cancellation through the full-duplex techniques.

\subsection{Transmit Signal Model}
Consider an OFDM frame with $L$ OFDM symbols. Let $f_c$ denote the carrier frequency, $M$ denote the number of subcarriers, $T_{\mathrm{s}}$ denote the elementary duration of an OFDM symbol, and $T_{\mathrm{cp}}$ denote the duration of the cyclic prefix (CP). Consequently, the subcarrier spacing, the overall bandwidth, and the overall symbol duration of the OFDM system are $\Delta f = \frac{1}{T_{\mathrm{s}}}$, $B = M \Delta f$, and $T_{\mathrm{tot}} = T_{\mathrm{s}} + T_{\mathrm{cp}}$, respectively. Then, the baseband transmit signal over an OFDM frame can be expressed as \cite{sturm2011waveform}
\begin{equation} \label{transmit_signal}
    \bar{\mathbf{x}}(t) = \frac{1}{\sqrt{M}} \sum_{l = 0}^{L-1} \sum_{m=0}^{M-1} \bar{\mathbf{x}}_m(l) e^{j 2\pi \delta_{m} \Delta f t} \mathrm{rect}\left( \frac{t - l T_{\mathrm{tot}}}{T_{\mathrm{tot}}} \right),
\end{equation} 
where $\bar{\mathbf{x}}_m(l) \in \mathbb{C}^{N \times 1}$ denotes the data signal on the $m$-th subcarriers in the $l$-th OFDM sysmbol, $\mathrm{rect}(t)$ denotes the rectangular function which has a value of $1$ if $t \in [0,1]$ and $0$ otherwise, and $\delta_m = \frac{2m-M+1}{2}$. The covariance matrix of signal $\bar{\mathbf{x}}_m(l)$ is defined as $\bar{\mathbf{R}}_m \triangleq \mathbb{E}[ \bar{\mathbf{x}}_m(l) \bar{\mathbf{x}}_m^H(l)] \succeq 0$. In this paper, we consider the average power constraint for each subcarrier, which is given by $\mathrm{tr}(\bar{\mathbf{R}}_m) \le P_m,$ with $P_m$ being the transmit power budget for the $m$-th subcarrier.

\begin{figure}[t!]
    \centering
    \includegraphics[width=0.3\textwidth]{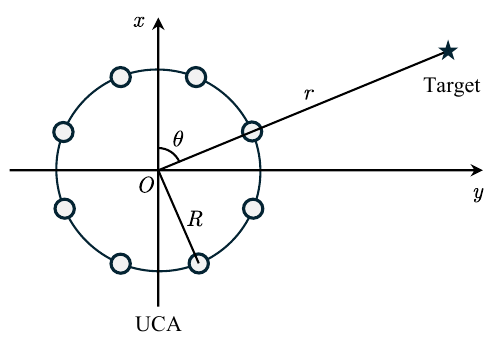}
    \caption{Geometry of the considered system.}
    \label{fig:system_model}
\end{figure}

\vspace{-0.1cm}
\subsection{Receive Signal Model}

We consider a two-dimensional coordinate system for the near-field sensing system, as illustrated in Fig. \ref{fig:system_model}. The origin of the coordinate system is put into the center of the UCA at the BS. For the target, let $r$ and $\theta$ denote its distance from the origin and the angle with respect to the $x$-axis. Then, the coordinate of the target can be expressed as $\mathbf{r} = [r \cos \theta, r \sin \theta ]^T$.
Let $d$ denote the antenna spacing of the UCA at the BS. The radius of the UCA is thus given by $R = \frac{N d}{2 \pi}$, which satisfies $R \le r$. In this case, the coordinate $\mathbf{s}_n$ of the $n$-th antenna can be expressed as $\mathbf{s}_n = \left[ R \cos \psi_n, R \sin \psi_n  \right]^T$,
where $\psi_n = \frac{2 \pi n}{N}$. The near-field propagation distance of the signal from the $n$-th antenna to the target is given by \cite{liu2023near_tutorial}
\begin{equation}
    r_n = \|\mathbf{r} -\mathbf{s}_n\| = \sqrt{ r^2 + R^2 - 2 r R \cos( \theta - \psi_n) }.
\end{equation}

Given the propagation distance, the round-trip propagation delay $\tau_{n,i}$  of the echo signal from the $n$-th antenna to the $i$-th antenna at the BS is $\tau_{n,i} = (r_n + r_i)/c$, where $c$ denotes the speed of light. Let $x_{m,n}(l) = [\mathbf{x}_m(l)]_n$ denote the baseband signals transmitted by the $n$-th antenna at the $m$-th subcarrier, the noiseless continuous-time baseband signal received at the $i$-th antenna at the BS can be modelled as \cite{goldsmith2005wireless}
\begin{align}
    &y_i(t) = \frac{1}{\sqrt{M}} \sum_{n=1}^{N} \sum_{l = 0}^{L-1} \sum_{m=0}^{M-1} \beta_{m, n, i} x_{m,n}(l) e^{j 2\pi \delta_{m} \Delta f \left(t - \tau_{n,i}\right)} \nonumber \\
    &\hspace{2.4cm} \times e^{-j 2 \pi f_c \tau_{n,i}} \mathrm{rect} \left( \frac{t - \tau_{n,i} - l T_{\mathrm{tot}}}{T_{\mathrm{tot}}} \right),
\end{align} 
where $\beta_{m,n,i}$ denote the frequency-dependent channel gain at the $m$-th subcarrier for the path from the $n$-th transmit antenna to the $i$-th receive antenna. According to the radar range equation \cite{richards2010principles}, $\beta_{m,n,i}$  can be modelled as 
\begin{equation}
    \beta_{m,n,i} = \frac{\sqrt{\epsilon_m} \beta_r}{r_n r_i}, \quad \epsilon_m = \frac{G_{t,m} G_{r,m} \lambda_m^2 }{(4 \pi)^3 }.
\end{equation}
In the above formula, $\beta_r$ models the amplitude and phase changes caused by the reflection at the target, $G_{t,m}$ and $G_{r,m}$ denote the antenna gain at frequency $f_m$, $\lambda_m = c/f_m$ denote the wavelength. We assume that the target is beyond the uniform-power distance, resulting in $r \approx r_1 \approx \dots \approx r_N$. In this circumstance, we have $\beta_{m,n,i}\approx \frac{\sqrt{\epsilon_m} \beta_r}{r^2}$. Then, the discrete-time signal model in the $l$-th OFDM symbol after removing CP can be obtained by the sampling $y_i(t)$ at time $t = l T_{\mathrm{tot}} + T_{\mathrm{cp}} + k \frac{T_s}{M}$ for $k = 0, \dots, M-1$. By omitting the constant terms, the noiseless discrete-time signal model is given by \cite{sturm2011waveform}:
\begin{align}
    y_{k,i}(l) = \frac{1}{\sqrt{M}} \sum_{n=1}^{N} \sum_{m=0}^{M-1} \frac{\sqrt{\epsilon_m} \beta_r}{r^2} x_{m,n}(l) e^{-j k_m (r_n + r_i) } e^{j 2\pi \frac{mk}{M}},
\end{align}  
where $k_m = 2\pi f_m/c$ is the wavenumber, $f_m = f_c + \delta_m \Delta f$ is the frequency of the $m$-th subcarrier. Let $\mathbf{y}_k(l) = [ y_{k,1}(l),\dots,y_{k,N}(l)  ]^T$ denote the vector collecting all signals received at the BS, which can be expressed as 
\begin{equation}
    \mathbf{y}_k(l) = \frac{1}{\sqrt{M}} \sum_{m=0}^{M-1}  \frac{\sqrt{\epsilon_m} \beta_r}{r^2} \mathbf{a}_m(\theta, r)\mathbf{a}_m^T(\theta, r) \bar{\mathbf{x}}_m(l) e^{j 2 \pi \frac{mk}{M}},
\end{equation}
where $\mathbf{a}_m(\theta, r)$ denotes the near-field array response vector and is given by 
\begin{equation}
    \mathbf{a}_m(\theta, r) = \left[ e^{-j k_m r_1},e^{-j k_m r_2},\dots,e^{-j k_m r_N} \right]^T,
\end{equation}
with $k_m = 2\pi f_m/c$ denoting the wavenumber. Then, the noisy signal received on the $m$-th subcarrier in the $l$-th OFDM symbol can be obtained by discrete Fourier transform (DFT) as follows 
\vspace{-0.2cm}
\begin{align}
    \mathbf{y}_m(l) = &\frac{1}{\sqrt{M}} \sum_{k=0}^{M-1} \mathbf{y}_k(l) e^{-j 2 \pi \frac{m k}{M}} + \mathbf{z}_m(l) \nonumber \\
    =  &\beta \mathbf{a}_m(\theta, r)\mathbf{a}_m^T(\theta, r) \mathbf{x}_m(l) + \mathbf{z}_m(l),
\end{align}  
where $\beta = \beta_r/r^2$ is the unknown channel gain, $\mathbf{x}_m(l) = \sqrt{\epsilon_m} \bar{\mathbf{x}}_m(l)$ is the equivalent transmit signal, and $\mathbf{z}_m(l)$ denote the additive white Gaussian noise with each entry obeying i.i.d. $\mathcal{CN}(0, \sigma_w^2)$. Aggregating $\mathbf{y}_{l,m}$ over $L$ OFDM symbols yields  
\begin{equation} \label{sensing_signal}
    \mathbf{Y}_m = [\mathbf{y}_m(1),\dots,\mathbf{y}_m(L) ] = \beta \mathbf{A}_m(\theta, r) \mathbf{X}_m + \mathbf{Z}_m,
\end{equation}
where $\mathbf{A}_m(\theta, r) = \mathbf{a}_m(\theta, r) \mathbf{a}_m^T( \theta, r)$, $\mathbf{X}_m = [\mathbf{x}_m(1),\dots,$ $\mathbf{x}_m(L)]$, and $\mathbf{Z}_m = [\mathbf{z}_m(1),\dots,\mathbf{z}_m(L)]$.
In the mono-static sensing setup, the data signal $\mathbf{X}_m$ is known at the BS. We assume a practical case where the prior knowledge of the target location is unavailable. In this case, to guarantee an optimal worst-case performance of sensing, the transmit signal need to be spatially white \cite{stoica2007probing}, i.e., 
\begin{equation} \label{isotropic_beam}
    \mathbf{R}_m \triangleq \mathbb{E}[ \mathbf{x}_m(l) \mathbf{x}_m^H(l) ] = \frac{\epsilon_m P_m}{N} \mathbf{I}_N, \forall m,
\end{equation}
where is referred to as isotropic beamforming method.
To guarantee the effective usage of each subcarrier, we assume that $\epsilon_m P_m$ has a constant value of $P$.

The problem for near-field sensing is to estimate the remaining unknown parameters, i.e., complex channel gain $\beta$, distance $r$, and angle $\theta$, related to the targets from the receive signals $\{\mathbf{Y}_m\}_{m=0}^{M-1}$ in \eqref{sensing_signal} based on the knowledge of $\mathbf{X}_m$. We focus primarily on the estimation of $r$ and $\theta$, i.e., the location information of the target.

\section{Performance Bounds and Analysis} \label{sec:analysis}
In this section, the performance bounds for angle and distance estimation are characterized and analyzed. In particular, the most popular CRB is considered, which provides a tight lower bound of mean-squared error for unbiased estimators under some general and mild conditions \cite{el2010conditional, 9439203, wang2023near, wang2023cram}.

\subsection{Cramér-Rao Bound}
We now derive the CRBs for estimating $\theta$ and $r$ from the signals $\{\mathbf{Y}_m\}_{m=0}^{M-1}$. To this end, we first stack the signals $\{\mathbf{Y}_m\}_{m=0}^{M-1}$ a single vector as follows:
\begin{align}
    \mathbf{y} = \underbrace{\begin{bmatrix}
        \mathrm{vec}(\beta \mathbf{A}_0(\theta, r) \mathbf{X}_0) \\
        \vdots \\
        \mathrm{vec}(\beta \mathbf{A}_{M-1}(\theta, r) \mathbf{X}_{M-1})
    \end{bmatrix}}_{\mathbf{u}} + \begin{bmatrix}
        \mathrm{vec}(\mathbf{Z}_0) \\
        \vdots \\
        \mathrm{vec}(\mathbf{Z}_{M-1})
    \end{bmatrix}.
\end{align}
Following the results in \cite{wang2023cram}, we define vectors $\mathbf{u}_{\theta} = \frac{\partial \mathbf{u}}{\partial \theta}$ and $\mathbf{u}_{r} = \frac{\partial \mathbf{u}}{\partial r}$, and the following matrix 
\begin{equation} \label{Q_matrix}
    \mathbf{Q} = \begin{bmatrix}
        \|\mathbf{u}_{\theta} \|^2 \sin^2 \Omega & \frac{\Re\{ \mathbf{u}^H \mathbf{\Phi} \mathbf{u} \}}{\|\mathbf{u}\|^2} \\
        \frac{\Re\{ \mathbf{u}^H \mathbf{\Phi} \mathbf{u} \}}{\|\mathbf{u}\|^2} &  \|\mathbf{u}_{r} \|^2 \sin^2 \Theta
    \end{bmatrix},
\end{equation}
where $\sin^2 \Omega = 1 - \frac{|\mathbf{u}_{\theta}^H \mathbf{u}|^2}{ \|\mathbf{u}_{\theta}\|^2 \|\mathbf{u}\|^2 }$, $\sin^2 \Theta = 1 - \frac{|\mathbf{u}_{r}^H \mathbf{u}|^2}{ \|\mathbf{u}_{r}\|^2 \|\mathbf{u}\|^2 }$, and $\mathbf{\Phi} = \mathbf{u}_{\theta}^H \mathbf{u}_r \mathbf{I} - \mathbf{u}_{\theta} \mathbf{u}_r^H$. Then, the CRBs for estimating $\theta$ and $r$ can be respectively calculated by \cite{ wang2023cram}
\begin{equation}
    \label{CRB_a}
    \mathrm{CRB}_{\theta} = \frac{\sigma_w^2  \|\mathbf{u}_{r} \|^2 \sin^2 \Theta}{2 \det \mathbf{Q}}, \quad \mathrm{CRB}_r = \frac{\sigma_w^2  \|\mathbf{u}_{\theta} \|^2 \sin^2 \Omega}{2 \det \mathbf{Q}}.
\end{equation}   
It can be observed that the value of CRBs is determined by the terms $\|\mathbf{u}\|^2$, $\|\mathbf{u}_{\theta}\|^2$, $\|\mathbf{u}_{r}\|^2$, $\mathbf{u}_{\theta}^H \mathbf{u}_r$, $\mathbf{u}_{\theta}^H \mathbf{u}$, and $\mathbf{u}_{r}^H \mathbf{u}$. Specifically, the expression of $\|\mathbf{u}\|^2$ is given by 
\begin{align} \label{eqn_calculate_u}
    \|\mathbf{u}\|^2 = & \sum_{m=0}^{M-1} \|\mathrm{vec}(\beta \mathbf{A}_m(\theta, r) \mathbf{X}_m)\|^2 \nonumber \\ 
    \overset{(a)}{\approx} &\sum_{m=0}^{M-1} \frac{|\beta|^2 P L}{N} \mathrm{tr} \left( \mathbf{A}_m(\theta, r) \mathbf{A}^H_m(\theta, r) \right) = |\beta|^2 P LNM,
\end{align}  
where $(a)$ stems from the equality $\|\mathrm{vec}(\mathbf{X})\|^2 = \mathrm{tr}(\mathbf{X} \mathbf{X}^H)$ and the approximation $\frac{1}{L} \mathbf{X}_m \mathbf{X}_m^H$ $\approx \mathbb{E}[ \mathbf{x}_m(l) \mathbf{x}_m(l)^H ] = \frac{P}{N}\mathbf{I}_N$. 
Furthermore, to calculate the intermediate parameters involving $\mathbf{u}_{\theta}$ and $\mathbf{u}_r$, we first define the following derivatives:
\begin{align}
    \dot{\mathbf{G}}_{\theta,m} \triangleq &\frac{\partial \mathbf{A}_m(\theta, r)}{\partial \theta} = -j k_m (\mathbf{\Theta} \mathbf{A}_m(\theta, r) + \mathbf{A}_m(\theta, r) \mathbf{\Theta}), \\
    \dot{\mathbf{G}}_{r,m} \triangleq &\frac{\partial \mathbf{A}_m(\theta, r)}{\partial r} = -j k_m (\mathbf{\Upsilon} \mathbf{A}_m(\theta, r) + \mathbf{A}_m(\theta, r) \mathbf{\Upsilon}),
\end{align}
where $\mathbf{\Theta}$ and $\mathbf{\Upsilon}$ are diagonal matrices whose $n$-th diagonal entries are $[\mathbf{\Theta}]_{n,n} = \frac{\partial r_n}{\partial \theta}$ and $[\mathbf{\Upsilon}]_{n,n} = \frac{\partial r_n}{\partial r}$, respectively.   
Then, $\mathbf{u}_{\theta}$ and $\mathbf{u}_r$ can be reformulated as 
\begin{equation}
    \mathbf{u}_i = \begin{bmatrix}
        \mathrm{vec}(\beta \dot{\mathbf{G}}_{i,0} \mathbf{X}_0) \\
        \vdots \\
        \mathrm{vec}(\beta \dot{\mathbf{G}}_{i,M-1} \mathbf{X}_{M-1})
    \end{bmatrix}, \forall i \in \{\theta, r\}.
\end{equation}  
Then, following the similar process as \eqref{eqn_calculate_u}, the expressions of $\|\mathbf{u}_{\theta}\|^2$ is given by 
\begin{align} \label{eq_22}
    \|\mathbf{u}_{\theta}\|^2 = &\sum_{m=0}^{M-1}  \frac{|\beta|^2 k_m^2 P L}{N} \left\|\mathbf{\Theta} \mathbf{A}_m(\theta, r) + \mathbf{A}_m(\theta, r) \mathbf{\Theta} \right\|_F^2 \nonumber \\ = & \frac{2 |\beta|^2 k_0^2 P L \tilde{M}}{N} \left( N u_{\theta} + c_{\theta}^2 \right),
\end{align} 
where $k_0 = 2\pi/c$, $\tilde{M} = \sum_{m=0}^{M-1} f_m^2 = M f_c^2 + \frac{M(M^2-1)}{12} \Delta f^2$, $u_{\theta} = \sum_{n=1}^N \left( \frac{\partial r_n}{\partial \theta} \right)^2$, and $c_{\theta} = \sum_{n=1}^N \frac{\partial r_n}{\partial \theta}$. 
Similarly, it can be shown that 
\begin{align}
    &\|\mathbf{u}_r\|^2 = \frac{2 |\beta|^2 k_0^2 P L \tilde{M}}{N} \left( N u_r + c_r^2 \right), \\
    &\mathbf{u}_{\theta}^H \mathbf{u}_r  = \frac{2 |\beta|^2 k_0^2 P L \tilde{M}}{N} \left( N \eta + c_{\theta} c_r \right),\\
    &\mathbf{u}_{\theta}^H \mathbf{u} = -2 j |\beta|^2 k_0 P L \bar{M} c_{\theta}, \\  
    \label{eq_24}
    &\mathbf{u}_r^H \mathbf{u} = -2 j |\beta|^2 k_0 P L \bar{M} c_r,
\end{align}
where $u_r = \sum_{n=1}^N \left( \frac{\partial r_n}{\partial r} \right)^2$, $c_r = \sum_{n=1}^N \frac{\partial r_n}{\partial r}$, $\eta = \sum_{n=1}^N  \frac{\partial r_n}{\partial \theta} \frac{\partial r_n}{\partial r}$, and $\bar{M} = \sum_{m=0}^{M-1} f_m = M f_c$.
To obtain the closed-form expressions of the CRBs, we first derive the following lemma. 
\begin{lemma}
    \emph{
        If $N \gg 1$ and $R \le r$, the closed-form expressions of $u_{\theta}$, $u_r$, $c_{\theta}$, $c_r$, and $\eta$ can be derived as   
        \begin{align}
            &u_{\theta} = \frac{R^2 N}{2}, u_r = N - \frac{R^2 N}{2 r^2}, \\ &c_r = N K\left( \frac{r}{R} \right), c_{\theta} = 0, \eta = 0,
        \end{align}   
        where $K(\alpha)$ is a transcendental function given by 
        \begin{equation}
            K(\alpha) = \int_{0}^{2 \pi} \frac{\alpha - \cos x}{2\pi \sqrt{ 1 - 2 \alpha \cos x + \alpha^2 }} dx.
        \end{equation} 
    }
\end{lemma}

\begin{IEEEproof}
    Please refer to the Appendix.
\end{IEEEproof}

Based on the results in \textbf{Lemma 1}, the closed-form expression of the intermediate parameters in \eqref{eq_22}-\eqref{eq_24} can be obtained. By substituting them into \eqref{CRB_a}, the closed-form CRBs can be obtained, given in the following theorem.

\begin{theorem}
    \emph{
        The closed-form CRBs achieved by UCAs when $R \le r$ are given by 
        \begin{align} \label{cf_crb_t}
            \mathrm{CRB}_{\theta} = &\frac{6}{ \rho  L N M R^2 \left( 12 f_c^2 + B^2 - \Delta f^2  \right)}, \\
            \label{cf_crb_r}
            \mathrm{CRB}_r = &\frac{3}{\left( \splitdfrac{\rho L N M \Big[ 12 f_c^2 \big( 1 - \frac{R^2}{2r^2} - K^2\left( \frac{r}{R} \right) \big)}{ + \left(B^2 - \Delta f^2\right) \left( 1 - \frac{R^2}{2r^2} + K^2\left( \frac{r}{R} \right) \right) \Big] } \right)},
        \end{align}
        where $\rho = k_0^2 |\beta|^2 P/\sigma_w^2$, $R = N d/(2 \pi)$ is the radius (half of the aperture) of the UCA, and $B = M \Delta f$ is the signal bandwidth. 
    }
\end{theorem}

\begin{figure}[t!]
    \centering
    \includegraphics[width=0.5\textwidth]{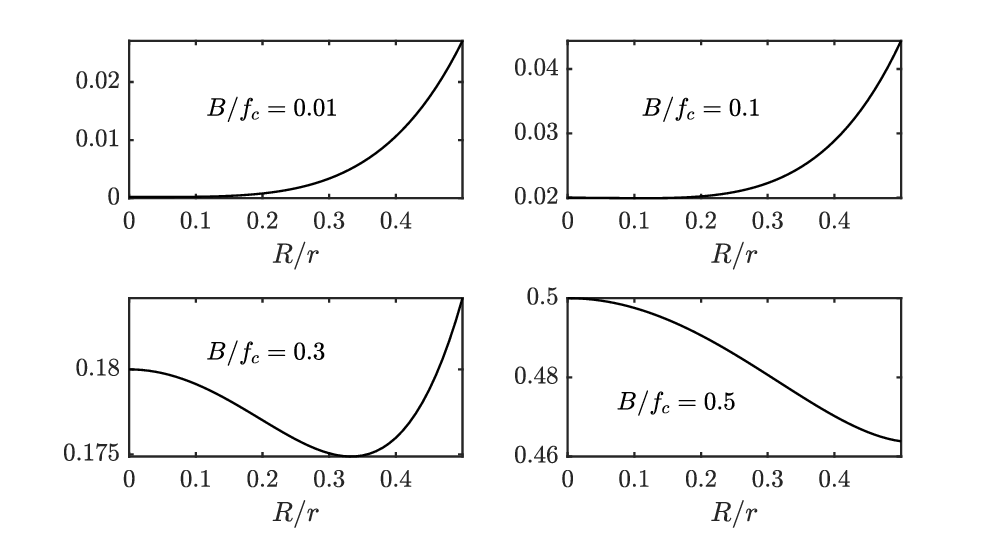}
    \caption{The numerical results of function $\Xi(\frac{R}{r}, \frac{B}{f_c})$.}
    \label{fig_K}
\end{figure} 

In \textbf{Theorem 1}, it can be observed that $\mathrm{CRB}_r$ is inversely proportional to the following function:
\begin{align} \label{K_func}
    &\Xi\left(\frac{R}{r}, \frac{B}{f_c}\right) = 12 \left( 1 - \frac{R^2}{2r^2} - K^2 \left( \frac{r}{R} \right)  \right) \nonumber \\
    &\hspace{2cm} + \frac{B^2 - \Delta f^2}{f_c^2} \left( 1 - \frac{R^2}{2 r^2} + K^2 \left( \frac{r}{R} \right)\right).
\end{align}
The behavior of this function when $M \gg 1$, i.e., $B^2 - \Delta f^2 \approx B^2$ is illustrated in Fig. \ref{fig_K}. From the closed-form expressions of CRBs in \textbf{Theorem 1} and the results in Fig. \ref{fig_K}, we notice the following (keeping in mind $R \le r$, $N \gg 1$, and $M \gg 1$).

\begin{itemize}

\item \textbf{Robustness of UCA:} $\mathrm{CRB}_{\theta}$ and $\mathrm{CRB}_r$ achieved by UCA is independent of the angle of the target $\theta$. This is fundamentally different from the conventional ULAs with an angular-dependent performance \cite{el2010conditional, 9439203, wang2023near, wang2023cram}. Therefore, UCAs can provide more stable performance than ULAs.

\item \textbf{Impact of Bandwidth:} Both $\mathrm{CRB}_{\theta}$ and $\mathrm{CRB}_r$ are $O(1/M)$ and decrease with both larger carrier frequency $f_c$ and bandwidth $B$. Compared to $\mathrm{CRB}_r$, $\mathrm{CRB}_{\theta}$ is less affected by the bandwidth $B$. This is because the $(12 f_c^2 + B^2 - \Delta f^2)$ term in its denominator is mainly affected by $12 f_c^2$ unless $B^2 - \Delta f^2$ has a comparable value to $12f_c^2$, which is impossible in practice. In contrast, $\mathrm{CRB}_r$ is significantly affected by the bandwidth $B$, which will be detailed in the sequel.  

\item \textbf{Impact of Array Size:} $\mathrm{CRB}_{\theta}$ is $O(1/N)$ and $O(1/R^2)$ but is independent of $r$. This suggests that increasing the array aperture is more advantageous for angle estimation than increasing the number of antennas. If the antenna spacing $d$ remains constant, the radius $R$ is proportional to the number of antennas, where $R = \frac{N d}{2 \pi}$. In this case, $\mathrm{CRB}_{\theta}$ is $O(1/N^3)$. On the contrary, $\mathrm{CRB}_r$ is $O(1/N)$ but exhibits a more complex dependence $R$ and $r$, characterized by the function $\Xi(\frac{R}{r}, \frac{B}{f_c})$ in \eqref{K_func}. According to the results in Fig. \ref{fig_K}, larger array apertures or closer targets (i.e., a larger ratio of $R/r$) generally enhance the performance of distance estimation with the practical ratio $B/f_c$ less than $0.1$. However, this trend appears to reverse at ultra-high values of $B/f_c$, exceeding $0.3$, where larger apertures or closer targets may actually degrade the performance of distance estimation. It is important to note that such high values of $B/f_c$ are generally uncommon in practice.

\item \textbf{Origins of Performance Gain:} In the considered system, the total number of observations for target sensing is $LMN$, where $L$, $M$, and $N$ denote the number of observations collected across the time, frequency, and space domains, respectively. Recall that with fixed $G$, $R$, and $B$, both $\mathrm{CRB}_{\theta}$ and $\mathrm{CRB}_r$ are $O(1/N)$ and $O(1/M)$. This suggests that when the array aperture and signal bandwidth remain constant, the performance gain from adding more antennas and subcarriers stems solely from the increase in the total number of observations, rather than changing the near-field and wideband effects. Conversely, modifications to the array aperture or signal bandwidth directly impact these effects, thereby fundamentally altering sensing performance.    
\end{itemize}  

\subsection{Asymptotic CRBs}

In this subsection, we dive further into the asymptotic behavior of CRBs to obtain more insights. Based on the \textbf{Theorem 1}, it is easy to prove the following corollaries.

\begin{corollary}
    \emph{As $r \rightarrow \infty$, the asymptotic $\mathrm{CRB}_r$ satisfies 
    \begin{equation}
        \label{cf_crb_r_r}
        \lim_{r \rightarrow +\infty} \mathrm{CRB}_r = \frac{3}{2\rho L N M \left(B^2 - \Delta f^2\right)}.
    \end{equation} 
    }
\end{corollary}

\begin{IEEEproof}
    As $r \rightarrow \infty$, we have $\lim_{r \rightarrow \infty} \frac{R}{2 r^2} = 0$ and 
    \begin{align}
        \lim_{r \rightarrow \infty} K \left(\frac{r}{R}\right) = \lim_{\alpha \rightarrow \infty} K \left(\alpha\right) = \int_{0}^{2 \pi} \frac{1}{2\pi} dx = 1.
    \end{align}
    By substituting these two limits into \eqref{cf_crb_r}, the expression in \eqref{cf_crb_r_r} can be obtained.
\end{IEEEproof}

\begin{figure}[t!]
    \centering
    \includegraphics[width=0.35\textwidth]{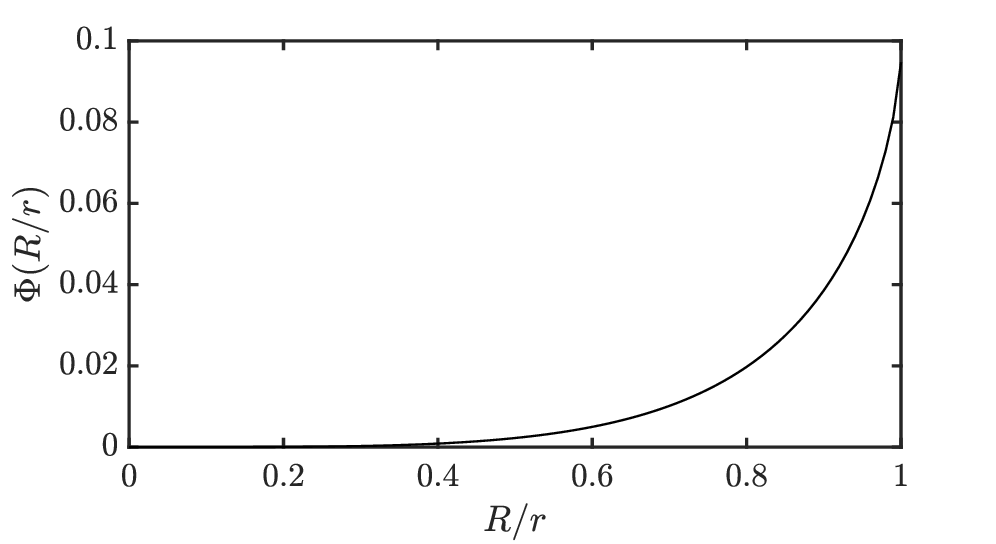}
    \caption{The numerical results of function $\Phi\left(\frac{R}{r}\right)$.}
    \label{fig_K_2}
\end{figure} 

\vspace{-0.05cm}
\begin{corollary}
    \emph{As $M \rightarrow 1$, the asymptotic $\mathrm{CRB}_r$ satisfies 
    \begin{equation} \label{cf_crb_m}
        \lim_{M \rightarrow 1} \mathrm{CRB}_r = \frac{1}{ 4 \rho L N f_c^2 \left( 1 - \frac{R^2}{2r^2} - K^2\left(\frac{r}{R}\right) \right) }. 
    \end{equation} 
    }
\end{corollary}

\begin{IEEEproof}
    As $M \rightarrow 1$, we have $B = \Delta f$, thus $B^2 - \Delta f^2 = 0$. Substituting this result into \eqref{cf_crb_r} yields \eqref{cf_crb_m}.    
\end{IEEEproof}

\begin{corollary}
    As $r \rightarrow \infty$ and $M \rightarrow 1$, the asymptotic $\mathrm{CRB}_r$ satisfies
    \begin{equation}
        \lim_{r \rightarrow \infty, M \rightarrow 1} \mathrm{CRB}_r = \infty. 
    \end{equation}
\end{corollary}

\begin{IEEEproof}
    The above results can be readily obtained according to \textbf{Corollaries 1} and \textbf{2}.
\end{IEEEproof}

According to \textbf{Corollary 2}, in the single-carrier system, the $\mathrm{CRB}_r$ is inversely proportional to the following function:
\begin{equation}
    \Phi\left(\frac{R}{r}\right) = 1 - \frac{R^2}{2r^2} - K^2 \left(\frac{r}{R}\right),
\end{equation}
The behavior of this function is illustrated in Fig. \ref{fig_K_2}. From the asymptotic CRBs in \textbf{Corollaries 1-3} and the results in Fig. \ref{fig_K_2}, we notice the following additional insights.
\begin{itemize}
    \item \textbf{Corollary 1} presents $\mathrm{CRB}_r$ in \emph{far-field systems}, where $r \rightarrow \infty$. In far-field systems, when the number of observations $LNM$ remains constant, $\mathrm{CRB}_r$ is merely related to the bandwidth $B$. This finding aligns with previous research on far-field sensing \cite{guerci2015joint}.    
    \item \textbf{Corollary 2} describes $\mathrm{CRB}_r$ in \emph{single-carrier systems}. According to the results in Fig. \ref{fig_K_2}, a larger array aperture or a closer target (i.e., a larger ratio of $R/r$) can always lead to a better performance of distance estimation. This result is consistent with the existing studies on single-carrier near-field sensing \cite{el2010conditional, 9439203, wang2023near, wang2023cram}.
    \item \textbf{Corollary 3} explores $\mathrm{CRB}_r$ in \emph{single-carrier far-field systems}. In this scenario, we have $\mathrm{CRB}_r = \infty$. This suggests an unbounded estimation error, rendering distance estimation infeasible under these conditions.
\end{itemize}

\vspace{-0.1cm}
\section{Numerical Results} \label{sec:results}
In this section, numerical results are provided to validate the analytical results. Unless otherwise specified, we set $r = 15$ m, $\theta = 90^\circ$, $f_c = 30$ GHz, $B = 10$ MHz, $L = 256$, $M = 256$, $N = 256$, $R = 0.5$ m, and $|\beta|^2 P/\sigma_w^2 = 0$ dB.

\begin{figure}[t!]
    \centering
    \includegraphics[width=0.43\textwidth]{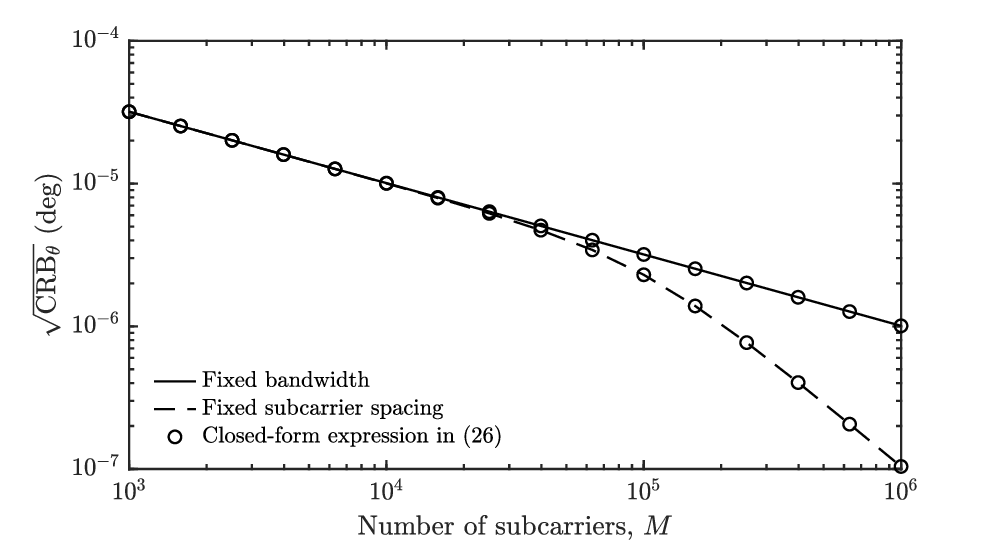}
    \includegraphics[width=0.43\textwidth]{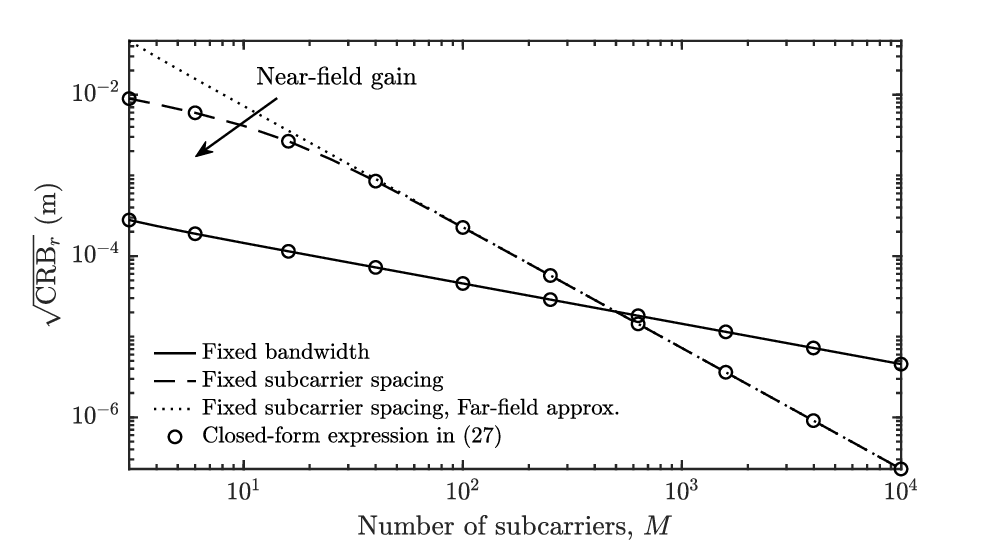}
    \caption{CRBs versus the number of subcarriers under the conditions of the fixed bandwidth of $B = 500$ MHz and the fixed subcarrier spacing of $1$ MHz, respectively.}
    \label{fig_CRB_M}
\end{figure}

Fig. \ref{fig_CRB_M} examines the influence of bandwidth on sensing performance under two different conditions: fixed bandwidth and fixed subcarrier spacing. In the case of fixed bandwidth, increasing the number of subcarriers only leads to more observation samples. Conversely, in the fixed subcarrier spacing scenario, increasing the number of subcarriers also expands the bandwidth. From Fig. \ref{fig_CRB_M}(a),  it can be observed that for angle estimation, increasing bandwidth has a marginal effect on its accuracy unless the bandwidth is extremely large, e.g., when $M \ge 10^5$ with $1$ MHz subcarrier spacing. In this case, the bandwidth is $B = 10^5 \times 1 \text{ MHz} = 100 \text{ GHz}$. This bandwidth is much larger than the carrier frequency $f_c$, which is impossible. Therefore, in practice system setup with $B \ll f_c$, bandwidth has negligible impact on angle estimation. Furthermore, for distance estimation, bandwidth has a much more significant impact, leading to a faster decrease in CRB. Furthermore, it reveals that the near-field gain in distance estimation is substantial when the bandwidth is small, but becomes negligible when the bandwidth is extremely large.

\begin{figure}[t!]
    \centering
    \includegraphics[width=0.43\textwidth]{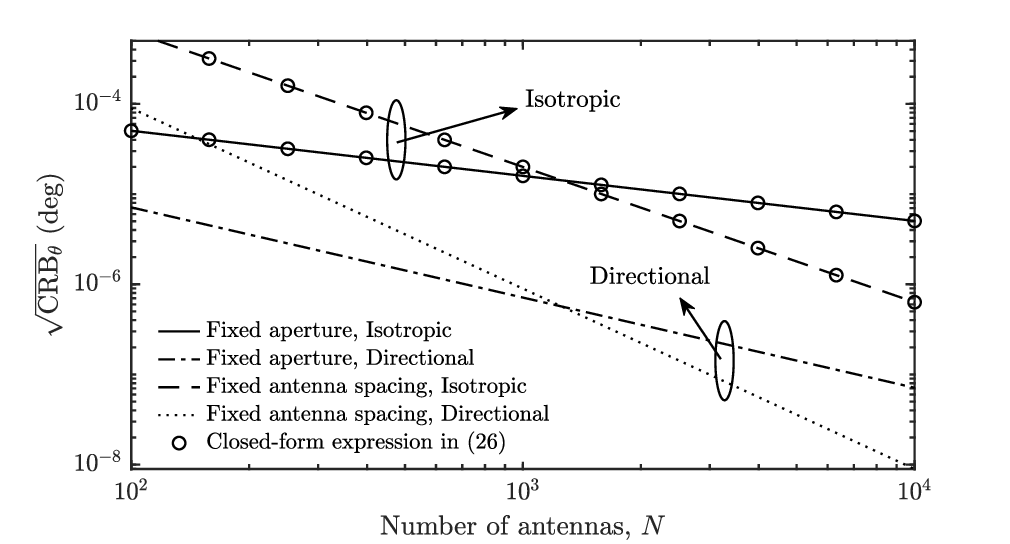}
    \includegraphics[width=0.43\textwidth]{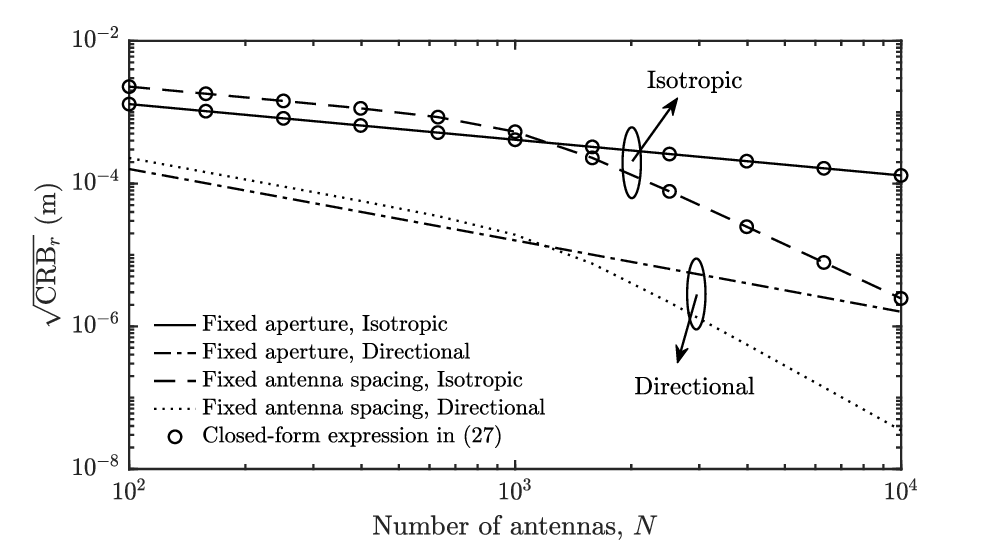}
    \caption{CRBs versus the number of antennas under the conditions of the fixed aperture of $R = 0.5$ m  and the fixed antenna spacing of $d = c/(2f_c)$, respectively.}
    \label{fig_CRB_N}
\end{figure}

Fig. \ref{fig_CRB_N} explores the impact of array size on sensing performance with either fixed aperture or fixed antenna spacing under different beamforming strategies. Apart from the spatially white BF in \eqref{isotropic_beam}, we also examine the directional beamforming that maximize the illumination power at the target location, leading to a transmit covariance matrix $\mathbf{R}_m = \frac{P}{N} \mathbf{a}_m^*(\theta, r) \mathbf{a}_m^T(\theta, r)$ \cite{stoica2007probing}. For fixed aperture, more antennas only lead to more observation samples. For fixed antenna spacing, more antennas also enlarge the array aperture, thus enhancing the near-field effect. It can be observed that the CRBs for angle and distance estimation exhibit similar trends as the number of antennas $N$ increases. First, both angle and distance estimation derive greater benefits from larger array apertures rather than merely increasing the number of antennas. Second, both metrics significantly improve with appropriate beamforming optimization.

\begin{figure}[t!]
    \centering
    \includegraphics[width=0.43\textwidth]{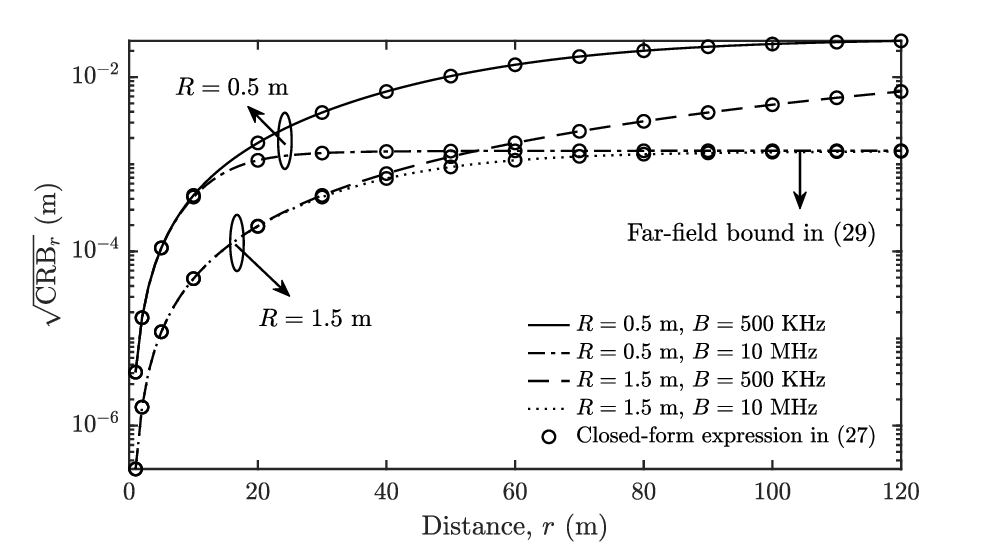}
    \caption{CRBs versus the target's distance under the conditions of different array apertures and signal bandwidths.}
    \label{fig_CRB_r}
\end{figure}

Fig. \ref{fig_CRB_r} explores the effect of target distance on the accuracy of distance estimation across different aperture and bandwidth configurations. There are two key observations. First, the accuracy of distance estimation approaches the far-field bound more rapidly with larger bandwidths, indicating that a larger bandwidth diminishes the extent of the near-field effect.  Second, when the target distance is relatively moderate (e.g., $r \le 60$ m), expanding the array aperture (without the addition of more antennas) is more advantageous than increasing the bandwidth, without incurring additional hardware costs or using more spectrum resources.

\vspace{-0.05cm}
\section{Conclusion} \label{sec:conclusion}
This paper studied the joint impact of bandwidth and array size on near-field sensing with circular arrays based on the CRB framework. The developed results suggested a highly coupled relationship between bandwidth and array size for the sensing performance. The developed results also included the existing results as special cases, providing a more accurate model for performance evaluation in practice. 


\section*{Appendix}
For UCAs, partial derivatives of the propagation distance $r_n$ with respect to $r$ and $\theta$ are given by
\begin{align}
    \frac{\partial r_n}{\partial \theta} = \frac{r R \sin(\theta - \frac{2\pi n}{N})}{\sqrt{ r^2 + R^2 - 2 r R \cos( \theta - \frac{2\pi n}{N}) }}, \\
    \frac{\partial r_n}{\partial r} = \frac{r - R \cos(\theta - \frac{2\pi n}{N})}{\sqrt{ r^2 + R^2 - 2 r R \cos( \theta - \frac{2\pi n}{N}) }}.
\end{align}
By defining $\delta = \frac{2\pi}{N}$, the parameter $\tilde{u}_{\theta}$ can be derived as     
\begin{align}
    \label{AD_1}
    \tilde{u}_\theta & = \sum_{n=1}^N \left( \frac{\partial r_n}{\partial \theta} \right)^2 = \sum_{n=1}^{N} \frac{r^2 R^2 \sin^2(\theta - \frac{2\pi n}{N})}{ r^2 + R^2 - 2 r R \cos( \theta - \frac{2\pi n}{N}) } \nonumber \\
    & = \frac{r^2 R^2}{\delta} \sum_{n=1}^{N} \frac{\sin^2(\theta - n \delta)}{r^2 + R^2 - 2 r R \cos(\theta - n \delta)} \delta  \nonumber \\
    &\overset{(a)}{\approx} \frac{r^2 R^2 N}{2\pi} \int_{0}^{2 \pi} \frac{\sin^2 x}{r^2 + R^2 - 2 r R \cos x} dx \nonumber \\
    & = \frac{r^2 R^2 N}{2 \pi} \bigg( \int_{0}^{\pi} \frac{\sin^2 x}{r^2 + R^2 - 2 r R \cos x} dx \nonumber \\
    &\hspace{1cm} + \int_{0}^{\pi} \frac{\sin^2 x}{r^2 + R^2 + 2 r R \cos x} dx \bigg) \overset{(b)}{=} \frac{R^2 N}{2},
\end{align}
where approximation $(a)$ is obtained based on $\delta \ll 1$ when $N \gg 1$ and step $(b)$ is derived based on the integral formula \cite[Eq. (3.613.3)]{gradshteyn2014table} when $R \le r$. Similarly, the remaining parameters can be derived as follows:    
\begin{align}
    \tilde{u}_r &= N - \frac{1}{r^2} \tilde{u}_{\theta} \approx N - \frac{R^2 N }{2 r^2},
\end{align}
\vspace{1cm}
\begin{align}
    \tilde{c}_{\theta} & \approx \int_{0}^{2 \pi} \frac{r R N \sin x}{2 \pi \sqrt{ r^2 + R^2 - 2r R_s \cos x }} dx \overset{(c)}{=} 0, \\
    \tilde{c}_r & \approx \int_{0}^{2\pi} \frac{N (r - R \cos x)}{ 2\pi \sqrt{r^2 + R^2 - 2 r R \cos x} } dx = N K\left(\frac{r}{R}\right), \\
    \label{AD_2}
    \tilde{\eta} &\approx \int_{0}^{2 \pi} \frac{ r R N (R \sin x \cos x -r \sin x)}{2\pi(r^2 + R^2 - 2 r R \cos x)} dx \overset{(d)}{=} 0,
\end{align}
where steps $(c)$ and $(d)$ are obtained according to the symmetry property of the functions and function $K(\alpha)$ is given by 
\begin{equation}
    K(\alpha) = \int_{0}^{2\pi} \frac{\alpha - \cos x}{ 2\pi \sqrt{1 - 2\alpha \cos x + \alpha^2}} dx.
\end{equation}
It can be proved that the function $K(\alpha)$ is a transcendental function that does not have a closed-form expression.
The proof of Lemma 1 is thus completed.

\bibliographystyle{IEEEtran}
\bibliography{reference/mybib}

\end{document}